\documentclass[12pt, a4paper]{article}

\usepackage{amsfonts,amsmath,amssymb,amsthm,latexsym}
\usepackage[all]{xy}

\newtheorem{proposition}{Proposition}[section]
\newtheorem{lemma}[proposition]{Lemma}

\theoremstyle{definition}
\newtheorem{definition}[proposition]{Definition}

\theoremstyle{remark}

\newcommand{\selabel}[1]{\label{se:#1}}
\newcommand{\seref}[1]{Section~\ref{se:#1}}

\newcommand{\prlabel}[1]{\label{pr:#1}}
\newcommand{\prref}[1]{Proposition~\ref{pr:#1}}

\newcommand{\delabel}[1]{\label{de:#1}}

\author{K. Kanakoglou, Physics Department, Aristotle University \\
of Thessaloniki, 54124, Thessaloniki, GREECE, \\
kanakoglou@hotmail.com \and C. Daskaloyannis, Mathematics
Department, Aristotle University \\
of Thessaloniki, 54124, Thessaloniki, GREECE, \\
 daskalo@math.auth.gr}
\title{Variants of bosonisation in Parabosonic algebra.
 The Hopf and super-Hopf structures.
 \footnote{J. of Phys. A: Math. and Theor.,
 v.\textbf{41}, 105203, (2008)}}

\date{July 2007}

\usepackage{appendix}

\begin{document}

\maketitle

\begin{abstract}
Parabosonic algebra in finite or infinite degrees of freedom is
considered as a $\mathbb{Z}_{2}$-graded associative algebra, and
is shown to be a $\mathbb{Z}_{2}$-graded (or: super) Hopf algebra.
The super-Hopf algebraic structure of the parabosonic algebra is
established directly without appealing to its relation to the
$osp(1/2n)$ Lie superalgebraic structure. The notion of super-Hopf
algebra is equivalently described as a Hopf algebra in the braided
monoidal category ${}_{\mathbb{CZ}_{2}}\mathcal{M}$. The
bosonisation technique for switching a Hopf algebra in the braided
monoidal category ${}_{H}\mathcal{M}$ (where $H$ is a
quasitriangular Hopf algebra) into an ordinary Hopf algebra is
reviewed. In this paper we prove that  for the parabosonic algebra
$P_{B}$, beyond the application of the bosonisation technique to
the original super-Hopf algebra, a bosonisation-like construction
is also achieved using two operators, related to the parabosonic
total number operator. Both techniques switch the same super-Hopf
algebra $P_{B}$ to an ordinary Hopf algebra, producing thus two
different variants of $P_{B}$, with ordinary Hopf structure.
\end{abstract}

\section{Introduction} \selabel{intro}
Parabosonic algebra has a long history both in theoretical and
mathematical physics. Although, formally introduced in the fifties
by Green \cite{Green}, Greenberg-Messiah \cite{GreeMe}, and Volkov
\cite{Vo} in the context of second quantization, its history
traces back to the fundamental conceptual problems of quantum
mechanics; in particular to Wigner's approach to first
quantization \cite{Wi}.  In quantum mechanics we consider a unital
associative non-commutative algebra, generated in terms of the
generators $p_{i}, q_{i}, I$, $i=1,\ldots,n$ and relations (we
have set $\hbar = 1$):
\begin{equation} \label{CCR}
[q_{i}, p_{j}] = i \delta_{ij} I, \qquad [q_{i},
q_{j}] = [p_{i}, p_{j}] = 0
\end{equation}
$I$ is of course the unity of the algebra and $[x, y]$ stands for
$xy-yx$. The states of the system are  vectors of a Hilbert space,
where the elements of the above mentioned algebra act. The
dynamics is determined by the Heisenberg equations of motion (we
have set $\hbar = 1$):
\begin{equation} \label{Heisenberg}
i \frac{dq_{i}}{dt} = [q_{i}, H],\qquad i
\frac{dp_{i}}{dt} = [p_{i}, H]
\end{equation}
 Relations \eqref{CCR} are known in the physical
literature as the Weyl algebra, or the Heisenberg-Weyl algebra or
more commonly as the Canonical Commutation Relations often
abbreviated as CCR. Their central importance for the quantization
procedure, lies in the fact that if one accepts the algebraic
relations \eqref{CCR} together with the quantum dynamical
equations \eqref{Heisenberg} then it is an easy matter (see
\cite{Ehrenf}) to extract the classical Hamiltonian equations of
motion while on the other hand the acceptance of the classical
Hamiltonian equations together with \eqref{CCR} reproduces the
quantum dynamics exactly as described by \eqref{Heisenberg}. We do
not consider arbitrary Hamiltonians of course but functions of the
form $H = \sum_{i=1}^{n}p_{i}^{2} + V(q_{1},\ldots,q_{n})$ which
however are general enough for simple physical systems. In this
way the CCR emerge as a fundamental link between the classical and
the quantum description of the dynamics.

 For  technical reasons it is common to use -instead of
the variables $p_{i}, q_{i}$- the linear combinations:
$$
b_{j}^{+} = \frac{1}{\sqrt{2}}(q_{j} - ip_{j}), \qquad
b_{j}^{-} = \frac{1}{\sqrt{2}}(q_{j} + ip_{j})
$$
for $j=1,\ldots,n$ in terms of which \eqref{CCR} become:
\begin{equation} \label{CCRbose}
[b_{i}^{-}, b_{j}^{+}] =  \delta_{ij} I, \qquad
[b_{i}^{-}, b_{j}^{-}] = [b_{i}^{+}, b_{j}^{+}] = 0
\end{equation}
for $i,j=1,\ldots,n$. These latter relations are usually called
the bosonic algebra (of n bosons), and in they case of the
infinite degrees of freedom $i,j = 1, 2, \ldots \ $ they become
the starting point of the free field theory (i.e.: second
quantisation).

In 1950 E.P. Wigner in a two page publication \cite{Wi}, noticed
that what the above approach implies is that the CCR \eqref{CCR}
are sufficient conditions -but not necessary- for the equivalence
between the classical Hamiltonian equations  and the Heisenberg
quantum dynamical equations \eqref{Heisenberg}. In a kind of
reversing the problem, Wigner posed the question of looking for
necessary conditions for the simultaneous fulfillment of classical
and quantum dynamical equations. Working with the simplest example
of a single, one dimensional harmonic oscillator, he stated an
infinite set of solutions for the above mentioned problem . It is
worth noting that a particular irreducible representation of the
CCR was included as one special case among Wigner's infinite
solutions.

A few years latter in 1953, Green in his celebrated paper
\cite{Green} introduced the parabosonic algebra (in possibly
infinite degrees of freedom), by means of generators and
relations:
\begin{equation} \label{CCRparabose}
\begin{array}{c}
  \big[ B_{m}^{-}, \{ B_{k}^{+}, B_{l}^{-} \} \big] = 2\delta_{km}B_{l}^{-}  \\
           \\
  \big[ B_{m}^{-}, \{ B_{k}^{-}, B_{l}^{-} \} \big]= 0 \\
           \\
  \big[ B_{m}^{+}, \{ B_{k}^{-}, B_{l}^{-} \} \big] = - 2\delta_{lm}B_{k}^{-} - 2\delta_{km}B_{l}^{-}\\
\end{array}
\end{equation}
$k,l,m = 1, 2, \ldots$ and $\{x, y \}$ stands for $xy+yx$. Green
was primarily interested in field theoretic implications of the
above mentioned algebra, in the sense that he considered it as an
alternative starting point for the second quantisation problem,
generalizing \eqref{CCRbose}. However, despite his original
motivation he was the first to realize -see also \cite{OhKa}- that
Wigner's infinite solutions were nothing else but inequivalent
irreducible representations of the parabosonic algebra
\eqref{CCRparabose}. (See also the discussion in \cite{Pal1}). \\

This paper consists logically of two parts. The first part
includes \seref{1}, \seref{2}, \seref{3}. The basic elements for
the structure parabosonic algebra are presented. In \seref{1} we
state the definition and derive basic properties of the
parabosonic algebra in infinite degrees of freedom. The
parabosonic algebra is considered to be a $\mathbb{Z}_2$-graded
associative algebra with an infinite set of (odd) generators
$B_{i}^{\pm}$ for $i = 1, 2, ... \ $. It's $\mathbb{Z}_2$-grading
is inherited by the natural $\mathbb{Z}_2$-grading of the tensor
algebra. The notions of $\mathbb{Z}_2$-graded algebra and
$\mathbb{Z}_{2}$-graded tensor products \cite{Che}, are discussed
as a special examples of the more general and modern notions of
$\mathbb{G}$-module algebras ($\mathbb{G}$: a finite abelian
group) and of braiding in monoidal categories \cite{Mon, Maj2,
Maj3}. In \seref{2} the notion of the super-Hopf algebra is
presented in connection with the non-trivial quasitriangular
structure of the $\mathbb{CZ}_{2}$ group Hopf algebra and the
braided monoidal category of it's representations
${}_{\mathbb{CZ}_{2}}\mathcal{M}$. The super-Hopf algebraic
structure of the parabosonic algebra is established, without
appealing to its Lie superalgebraic structure, and this is the
central result of this part of the paper. Let us remark here  In
\seref{3}, for the sake of completeness, well known results
regarding the Lie superalgebraic structure of the parabosonic
algebra in finite degrees of freedom are reviewed.

The second part of the paper consists of \seref{4}. We begin the
section with a review of the bosonisation technique for switching
a Hopf algebra $A$ in a braided monoidal category $\mathcal{C}$
into an ordinary Hopf algebra. Although we do not present the
method in it's full generality (see \cite{Maj1}), we give
sufficient details for its application in a much more general
class of problems than those involved in the ``super" or even in
the $\mathbb{G}$-graded ($\mathbb{G}$ finite and abelian) case: We
consider the case of a Hopf algebra in the braided monoidal
category ${}_{H}\mathcal{M}$ where $H$ is some quasitriangular
Hopf algebra, and explain in detail how we can construct an
ordinary Hopf algebra out of it. The construction is achieved by
means of a smash product algebra $A \star H$, and uses older
results \cite{Mo}, \cite{Ra}, which guarantee the compatibility
between the algebraic and the coalgebraic structure, in order for
a smash product to be a Hopf algebra. The construction is such
that the (braided) modules of the original (braided) Hopf algebra
$A$ and the (ordinary) modules of the ``bosonised" (ordinary) Hopf
algebra $A \star H$ are in a bijective correspondence, providing
thus an equivalence of categories. We apply the method in the case
of the parabosonic algebra, i.e. the case for which $H =
\mathbb{CZ}_{2}$ equipped with it's non-trivial quasitriangular
structure, producing a ``variant" of the parabosonic algebra. This
variant $P_{B} \star \mathbb{CZ}_{2}$, which we will denote by
$P_{B(g)}$, is a smash product Hopf algebra between the
parabosonic super-Hopf algebra $P_{B}$ and the group Hopf algebra
$\mathbb{CZ}_{2}$, and it is a Hopf algebra in the ordinary sense
(and not in the ``super" sense). We explicitly state the structure
maps (multiplication, comultiplication, counity and the antipode)
for the (ordinary) Hopf algebraic structure of $P_{B(g)}$. Finally
one more variant of the bosonisation for the parabosonic algebra
is presented, which achieves the same object with the bosonisation
technique. We construct an algebra $P_{B(K^{\pm})}$, which is a
little ``bigger" than the parabosonic algebra $P_{B}$ or it's
bosonised form $P_{B(g)}$ and we establish it's (ordinary) Hopf
algebraic structure. So we prove that the bosonisation technique
is not unique.

 In what follows, all vector spaces and algebras and all
tensor products will be considered over the field of complex
numbers. Whenever the symbol $i$ enters a formula in another place
than an index, it always denotes the imaginary unit $i^{2} = -1$.
Furthermore, whenever formulas from physics enter the text, we use
the traditional convention: $\hbar = m = \omega = 1$. Finally, the
Sweedler's notation for the comultiplication is freely used
throughout the text.

\section{Super-algebraic structure of Parabosons} \selabel{1}
The parabosonic algebra, was originally defined in terms of
generators and relations by Green \cite{Green} and
Greenberg-Messiah \cite{GreeMe}. We begin with restating their
definition, in a modern algebraic context.
 Let us consider the vector space $V_{X}$ freely
generated by the elements: $X_{i}^{+}, X_{j}^{-}$, $i,j=1, 2, ...
\ $. Let $T(V_{X})$ denote the tensor algebra of $V_{X}$:
$$
T(V_{X}) = \bigoplus _{n \geq 0} T^{n}(V_{X})
$$
where $T^{0}(V_{X}) = \mathbb{C}$, $ \ T^{1}(V_{X}) = V_{X}$ and
for $n \geq 2$: $T^{n}(V_{X}) = V_{X} \otimes ... \otimes V_{X}$
the $n$-th tensor power of $V_{X}$. It is well known \cite{Che}
that $ \ T(V_{X})$ is -up to isomorphism- the free algebra
generated by the elements $X_{i}^{+}$, $X_{j}^{-}$ ($i,j=1, 2, ...
\ $) of the basis of $V_{X}$ or equivalently the non-commutative
polynomial algebra generated over the indeterminates  $X_{i}^{+}$,
$X_{j}^{-}$ ($i,j=1, 2, ... \ $). In $T(V_{X})$ we consider the
two-sided ideal $I_{P_{B}}$, generated by the following elements:
\begin{equation} \label{eq:pbdef}
 \big[ \{ X_{i}^{\xi},  X_{j}^{\eta}\}, X_{k}^{\epsilon}  \big] -
 (\epsilon - \eta)\delta_{jk}X_{i}^{\xi} - (\epsilon - \xi)\delta_{ik}X_{j}^{\eta}
\end{equation}
respectively, for all values of $\xi, \eta, \epsilon = \pm 1$ and
$i,j=1, 2, ... \ $. $ \ I_{X}$ is the unity of the tensor algebra.
$[A, B]$ stands for $A \otimes B - B \otimes A$ and $ \{A, B \}$
stands for $A \otimes B + B \otimes A$, where $A$ and $B$ are
arbitrary elements of the tensor algebra $T(V_{X})$. We now have
the following:
\begin{definition} \delabel{parabosonsbosons}
The parabosonic algebra in $P_{B}$  is the quotient algebra of the
tensor algebra $T(V_{X})$ of $V_{X}$ with the ideal $I_{P_{B}}$:
$$
P_{B} = T(V_{X}) / I_{P_{B}}
$$
\end{definition}
We denote by $\pi_{P_{B}}: T(V_{X}) \rightarrow P_{B}$ the
canonical projection. The elements $X_{i}^{+}$, $ \ X_{j}^{-}$, $
\ I_{X}$, where $i,j=1, 2, ... \ $ and $I_{X}$ is the unity of the
tensor algebra, are the generators of the tensor algebra
$T(V_{X})$. The elements $\pi_{P_{B}}(X_{i}^{+}),
\pi_{P_{B}}(X_{j}^{-}), \pi_{P_{B}}(I_{X}) \ $, $ \ i,j=1,...$ are
a set of generators of the parabosonic algebra $P_{B}$, and they
will be denoted by $B_{i}^{+}, B_{j}^{-}, I$ for $i,j=1, 2, ...$
respectively, from now on. $\pi_{P_{B}}(I_{X}) = I$ is the unity
of the parabosonic algebra. The generators of the parabosonic algebra
satisfy equ. (\ref{CCRparabose}).

Based on the above definitions we prove now the following
proposition:
\begin{proposition} \prlabel{parabosonstobosons}
The parabosonic algebra $P_{B}$ is a $\mathbb{Z}_{2}$-graded
associative algebra with it's generators $B_{i}^{\pm}$ for $i,j=1,
2, ...$, being odd elements.
\end{proposition}
\begin{proof}
It is obvious that the tensor algebra $T(V_{X})$ is a
$\mathbb{Z}_{2}$-graded algebra with the monomials being
homogeneous elements. If $x$ is an arbitrary monomial of the
tensor algebra, the degree of $x$ is denoted by $|x|={\rm deg}\,x$.
Then $|x|={\rm deg}(x) = 0$, namely $x$ is an even element,
if it constitutes of an even number of factors (an even number of
generators of $T(V_{X})$) and $|x|={\rm deg}\,(x) = 1$, namely $x$ is an odd
element, if it constitutes of an odd number of factors (an odd
number of generators of $T(V_{X})$). The generators $X_{i}^{+},
X_{j}^{-} \ $, $ \ i,j=1,...,n$ are odd elements in the above
mentioned gradation.
 In view of the above description we can easily conclude that the
$\mathbb{Z}_{2}$-gradation of the tensor algebra is immediately
``transfered" to the algebra $P_{B}$. The ideal $I_{P_{B}}$ is an
homogeneous ideal of the tensor algebra, since it is generated by
homogeneous elements of $T(V_{X})$. Consequently, the projection
homomorphism $\pi_{P_{B}}$ is an homogeneous algebra map of degree
zero, or we can equivalently say that it is an even algebra
homomorphism.
\end{proof}

The rise of the theory of quasitriangular Hopf algebras from the
mid-80's \cite{Dri} and thereafter and especially the study and
abstraction of their representations (see: \cite{Maj2, Maj3},
\cite{Mon} and references therein), has provided us with a novel
understanding of the notion and the properties of
$\mathbb{G}$-graded algebras,
where $\mathbb{G}$ is a finite abelian group.
 We are restricting ourselves to the simplest case
where $\mathbb{G} = \mathbb{Z}_{2}$ and we denote by $\{1, g\}$
the elements of the $\mathbb{Z}_{2}$ group (written
multiplicatively). An algebra $A$ being a $\mathbb{Z}_{2}$-graded
algebra (in the physics literature the term superalgebra is also
of widespread use) is equivalent to saying that $A$ is a
$\mathbb{CZ}_{2}$-module algebra, via the $\mathbb{Z}_{2}$-action
determined by:
 \[ 1 \triangleright a = a \; \mbox{ and }\; g \triangleright a
= (-1)^{|a|}a
\]
for any $a$ homogeneous in $A$ and $\ |a| \ $ it's degree.

 What we
actually mean is that $A$, apart from being an algebra is also a
$\mathbb{CZ}_{2}$-module and at the same time the structure maps
of $A$ (i.e.: the multiplication and the unity map which embeds
the field into the center of the algebra) are
$\mathbb{CZ}_{2}$-module maps, which is nothing else but
homogeneous linear maps of degree $0$ (or: even linear maps).
Stated more generally, the $\mathbb{G}$-grading of $A$ can be
equivalently described in terms of a specific action of the finite
abelian group $\mathbb{G}$ on $A$, thus in terms of a specific
action of the $\mathbb{CG}$ group algebra on $A$. This is not
something new. In fact such ideas already appear in works such as
\cite{CohMon} and \cite{Stee}.

In ref \cite{Maj3}, \cite{Mon} the construction of the tensor
products of $\mathbb{G}$-graded objects, is presented as a
consequence of the quasitriangularity of the $\mathbb{CG}$ group
Hopf algebra (for $\mathbb{G}$ a finite abelian group, see
\cite{Scheu1}) or in other words: as a consequence of the braiding
of the monoidal category ${}_{\mathbb{CG}}\mathcal{M}$ (category
of $\mathbb{CG}$-modules).

It is well known that for any group $\mathbb{G}$, the group
algebra $\mathbb{CG}$ equipped with the maps:
$$
\begin{array}{ccccc}
  \Delta(z) = z \otimes z &  & \varepsilon(z) = 1 &  & S(z) = z^{-1} \\
\end{array}
$$
for any $z \in \mathbb{G}$, becomes a Hopf algebra. Focusing again
in the special case $\mathbb{G} = \mathbb{Z}_{2}$, the fact that
$A$ is a  $\mathbb{Z}_2$-graded algebra is equivalently described
by saying that $A$ is an algebra in the braided monoidal category
of $\mathbb{CZ}_{2}$-modules ${}_{\mathbb{CZ}_{2}}\mathcal{M}$. In
this case the braiding is induced by the non-trivial
quasitriangular structure of the $\mathbb{CZ}_{2}$ Hopf algebra
i.e. by the non-trivial $R$-matrix:
\begin{equation} \label{eq:nontrivRmatrcz2}
R_{Z_{2}} = \frac{1}{2}(1 \otimes 1 + 1 \otimes g + g \otimes 1 -
g \otimes g)
\end{equation}

 We digress here for a moment, to recall
that (see \cite{Maj2, Maj3} or \cite{Mon}) if $(H,R_{H})$ is a
quasitriangular Hopf algebra through the $R$-matrix $R_{H} = \sum
R_{H}^{(1)} \otimes R_{H}^{(2)}$, then the category of modules
${}_{H}\mathcal{M}$ is a braided monoidal category, where the
braiding is given by a natural family of isomorphisms $\Psi_{V,W}:
V \otimes W \cong W \otimes V$, given explicitly by:
\begin{equation} \label{eq:braid}
\Psi_{V,W}(v \otimes w) = \sum (R_{H}^{(2)} \vartriangleright w)
\otimes (R_{H}^{(1)} \vartriangleright v)
\end{equation}
for any $V,W \in obj({}_{H}\mathcal{M})$. By $v,w$ we denote any
 elements of $V,W$ respectively.   \\
Combining eq. \eqref{eq:nontrivRmatrcz2} and \eqref{eq:braid} we
immediately get the braiding in the
${}_{\mathbb{CZ}_{2}}\mathcal{M}$ category:
\begin{equation} \label{symmbraid}
\Psi_{V,W}(v \otimes w) = (-1)^{|v||w|} w \otimes v
\end{equation}
 This is  a symmetric braiding,
since
 \[\Psi_{V,W} \circ \Psi_{W,V} = Id\]
  so we actually have a
symmetric monoidal category ${}_{\mathbb{CZ}_{2}}\mathcal{M}$,
rather than a truly braided one.

 The really important thing about the existence of the braiding
\eqref{symmbraid} is that it provides us with an alternative way
of forming tensor products of $\mathbb{Z}_{2}$-graded algebras. If
$A$ and $B$ are superalgebras with multiplications:
 \[
 m_{A}: A
\otimes A \rightarrow A\;  \mbox{ and  }\; m_{B}: B \otimes B \rightarrow B
\]
respectively, then the super vector space $A \otimes B$ (with the
obvious $\mathbb{Z}_{2}$-gradation) is equipped with the
associative multiplication
\begin{equation} \label{braidedtenspr}
(m_{A} \otimes m_{B})(Id \otimes \Psi_{B,A} \otimes Id): A \otimes
B \otimes A \otimes B \longrightarrow A \otimes B
\end{equation}
 given equivalently by:
\[
(a \otimes b)(c \otimes d) = (-1)^{|b||c|}ac \otimes bd
\]
for $b,c$ homogeneous in $B, A$ respectively. The tensor product
becomes a superalgebra (or equivalently an algebra in the braided
monoidal category of $\mathbb{CZ}_{2}$-modules
${}_{\mathbb{CZ}_{2}}\mathcal{M}$) which we will denote: $A
\underline{\otimes} B$ and call the braided tensor product algebra
from now on.

Let us close this description with two important remarks. First,
we stress that in \eqref{braidedtenspr} both superalgebras $A$ and
$B$ are viewed as $\mathbb{CZ}_{2}$-modules and as such we have $B
\otimes A \cong A \otimes B$ through $b \otimes c \mapsto
(-1)^{|c||b|} c \otimes b$. Second we underline that the tensor
product \eqref{braidedtenspr} had been already known from the past
\cite{Che} but rather as a special possibility of forming tensor
products of superalgebras than as an example of the more general
conceptual framework of the braiding applicable not only to
superalgebras but to any $\mathbb{G}$-graded algebra ($\mathbb{G}$
a finite abelian group) as long as $\mathbb{CG}$ is equipped with
a non-trivial quasitriangular structure or equivalently
\cite{Mon}, \cite{Scheu1}, a bicharacter on $\mathbb{G}$ is given.

\section{Super-Hopf structure of Parabosons: a braided
group}\selabel{2}

The notion of $\mathbb{G}$-graded Hopf algebra, for $\mathbb{G}$ a
finite abelian group, is not a new one neither in physics nor in
mathematics. The idea appears already in the work of Milnor and
Moore \cite{MiMo}, where we actually have $\mathbb{Z}$-graded Hopf
algebras. On the other hand, universal enveloping algebras of Lie
superalgebras are widely used in physics and they are examples of
$\mathbb{Z}_{2}$-graded Hopf algebras (see for example \cite{Ko},
\cite{Scheu}). These structures are strongly resemblant of Hopf
algebras but they are not Hopf algebras at least in the ordinary
sense.

Restricting again to the simplest case where $\mathbb{G} =
\mathbb{Z}_{2}$ we briefly recall this idea: An algebra $A$ being
a $\mathbb{Z}_{2}$-graded Hopf algebra (or super-Hopf algebra)
means first of all that $A$ is a $\mathbb{Z}_{2}$-graded
associative algebra (or: superalgebra). We now consider the
braided tensor product algebra $A \underline{\otimes} A$. Then $A$
is equipped with a coproduct
\begin{equation} \label{braidedcom}
\underline{\Delta} : A \rightarrow A \underline{\otimes} A
\end{equation}
which is an superalgebra homomorphism from $A$ to the braided
tensor product algebra  $A \underline{\otimes} A$ :
$$
\underline{\Delta}(ab) = \sum (-1)^{|a_{2}||b_{1}|}a_{1}b_{1}
\otimes a_{2}b_{2} = \underline{\Delta}(a) \cdot
\underline{\Delta}(b)
$$
for any $a,b$ in $A$, with $\underline{\Delta}(a) = \sum a_{1}
\otimes a_{2}$, $\underline{\Delta}(b) = \sum b_{1} \otimes
b_{2}$, and $a_{2}$, $b_{1}$ homogeneous. We emphasize here that
this is exactly the central point of difference between the
``super" and the ``ordinary" Hopf algebraic structure: In an
ordinary Hopf algebra $H$ we should have a coproduct $\Delta : H
\rightarrow H \otimes H$ which should be an algebra homomorphism
from $H$ to the usual tensor product algebra $H \otimes H$.

 Similarly, $A$ is equipped with an antipode $\underline{S} : A
\rightarrow A$ which is not an algebra anti-homomorphism (as it
should be in an ordinary Hopf algebra) but a  superalgebra
anti-homomorphism (or: ``twisted" anti-homomorphism or: braided
anti-homomorphism) in the following sense (for any homogeneous
$a,b \in A$):
\begin{equation} \label{braidedanti}
\underline{S}(ab) = (-1)^{|a||b|}\underline{S}(b)\underline{S}(a)
\end{equation}
The rest of the axioms which complete the super-Hopf algebraic
structure (i.e.: coassociativity, counity property, and
compatibility with the antipode) have the same formal description
as in ordinary Hopf algebras.

 Once again, the abstraction of the representation theory of
quasitriangular Hopf algebras provides us with a language in which
the above description becomes much more compact: We simply say
that $A$ is a Hopf algebra in the braided monoidal category of
$\mathbb{CZ}_{2}$-modules ${}_{\mathbb{CZ}_{2}}\mathcal{M}$ or: a
braided group where the braiding is given in equation
\eqref{symmbraid}. What we actually mean is that $A$ is
simultaneously an algebra, a coalgebra and a
$\mathbb{CZ}_{2}$-module, while all the structure maps of $A$
(multiplication, comultiplication, unity, counity and the
antipode) are also $\mathbb{CZ}_{2}$-module maps and at the same
time the comultiplication $\underline{\Delta} : A \rightarrow A
\underline{\otimes} A$ and the counit are algebra morphisms in the
category ${}_{\mathbb{CZ}_{2}}\mathcal{M}$ (see also \cite{Maj2,
Maj3} or \cite{Mon} for a more detailed description).
\\
We proceed now to the proof of the following proposition which
establishes the super-Hopf algebraic structure of the parabosonic
algebra $P_{B}$:
\begin{proposition} \prlabel{superHopfPb}
The parabosonic algebra  equipped with the even linear maps
$\underline{\Delta}: P_{B} \rightarrow P_{B} \underline{\otimes}
P_{B} \ \ $, $\ \ \underline{S}: P_{B} \rightarrow P_{B} \ \ $, $\
\ \underline{\varepsilon}: P_{B} \rightarrow \mathbb{C} \ \ $,
determined by their values on the generators:
\begin{equation} \label{eq:HopfPB}
\begin{array}{ccccc}
  \underline{\Delta}(B_{i}^{\pm}) = 1 \otimes B_{i}^{\pm} + B_{i}^{\pm} \otimes 1 &
  & \underline{\varepsilon}(B_{i}^{\pm}) = 0  & & \underline{S}(B_{i}^{\pm}) = - B_{i}^{\pm} \\
\end{array}
\end{equation}
for $i = 1, 2, \ldots \ $, becomes a super-Hopf algebra.
\end{proposition}
\begin{proof}
Recall that by definition $P_{B} = T(V_{X}) / I_{P_{B}}$. Consider
the linear map: $$\underline{\Delta}^{T}: V_{X} \rightarrow P_{B}
\underline{\otimes} P_{B}$$ determined by it's values on the basis
elements specified by: $$\underline{\Delta}^{T}(X_{i}^{\pm}) = I
\otimes B_{i}^{\pm} + B_{i}^{\pm} \otimes I$$ By the universality
of the tensor algebra this map is uniquely extended to a
superalgebra homomorphism: $\underline{\Delta}^{T}: T(V_{X})
\rightarrow P_{B} \underline{\otimes} P_{B}$. After lengthly
algebraic calculations (see Appendix \ref{app}) we can prove that:
\begin{equation}
\underline{\Delta}^{T}(\big[ \{ X_{i}^{\xi},  X_{j}^{\eta}\},
X_{k}^{\epsilon}  \big] -
 (\epsilon - \eta)\delta_{jk}X_{i}^{\xi} - (\epsilon -
 \xi)\delta_{ik}X_{j}^{\eta})= 0
\label{eq:comultinsuperparab}
\end{equation}
for all values of $\xi, \eta, \epsilon = \pm 1$ and $i,j=1, 2, ...
\ $. This means that $I_{P_{B}} \subseteq
ker(\underline{\Delta}^{T}$), which in turn implies that
$\underline{\Delta}^{T}$ is uniquely extended to a superalgebra
homomorphism: $\underline{\Delta}: P_{B} \rightarrow P_{B}
\underline{\otimes} P_{B}$, according to the following
(commutative) diagram:
\begin{displaymath}
\xymatrix{T(V_{X}) \ar[rr]^{\underline{\Delta}^{T}}
\ar[dr]_{\pi_{P_{B}}} & &
P_{B} \underline{\otimes} P_{B} \\
 & P_{B} \ar@{.>}[ur]_{ \underline{\Delta}} & }
\end{displaymath}
with values on the generators determined by \eqref{eq:HopfPB}.

Proceeding the same way we construct the maps $\
\underline{\varepsilon} \ $, $\ \ \underline{S} \ $, as determined
in \eqref{eq:HopfPB}.

For the case of $\underline{\varepsilon}$, we start defining the
trivial zero map
\[
\underline{\varepsilon}^T\,:\, V_x\to \{0\}\in \mathbb{C}
\]
and we (uniquely) extend its definition to a superalgebra
homomorphism $ \ \underline{\varepsilon}: P_{B} \rightarrow
\mathbb{C} \ $ following the commutative diagram:
\begin{displaymath}
\xymatrix{T(V_{X}) \ar[rr]^{\underline{\varepsilon}^{T}}
\ar[dr]_{\pi_{P_{B}}} & &
\mathbb{C} \\
 & P_{B} \ar@{.>}[ur]_{ \ \underline{\varepsilon}} & }
\end{displaymath}
with values on the generators determined by \eqref{eq:HopfPB}.

In the case of the antipode $\underline{S}$ we need the notion of
the $\mathbb{Z}_{2}$-graded opposite algebra (or: opposite
superalgera) $P_{B}^{op}$, which is a superalgebra defined as
follows: $P_{B}^{op}$ has the same underlying super vector space
as $P_{B}$, but the multiplication is now defined as: $a \cdot b =
(-1)^{|a||b|}ba$, for all $a,b \in P_{B}$. (In the right hand
side, the product is of course the product of $P_{B}$). We start
by defining a linear map
$$\underline{S}^{T}: V_{X} \rightarrow P_{B}^{op}$$ determined by:
$$\underline{S}^{T}(X_{i}^{\pm}) = -B_{i}^{\pm}$$ This map is (uniquely)
extended to a superalgebra homomorphism: $\underline{S}^{T}:
T(V_{X}) \rightarrow P_{B}^{op}$. Now we can compute:
\begin{equation} \label{eq:antipodeinsuperparab}
\underline{S}^{T}(\big[ \{ X_{i}^{\xi},  X_{j}^{\eta}\},
X_{k}^{\epsilon}  \big] -
 (\epsilon - \eta)\delta_{jk}X_{i}^{\xi} - (\epsilon -
 \xi)\delta_{ik}X_{j}^{\eta}) = 0
\end{equation}
for all values of $\xi, \eta, \epsilon = \pm 1$ and $i,j=1, 2, ...
\ $. This means that $I_{P_{B}} \subseteq ker(\underline{S}^{T})$,
which in turn implies that $\underline{S}^{T}$ is uniquely
extended to a superalgebra homomorphism $\underline{S}: P_{B}
\rightarrow P_{B}^{op}$, according to the following commutative
diagram:
\begin{displaymath}
\xymatrix{T(V_{X}) \ar[rr]^{\underline{S}^{T}}
\ar[dr]_{\pi_{P_{B}}} & &
P_{B}^{op} \\
 & P_{B} \ar@{.>}[ur]_{ \underline{S}} & }
\end{displaymath}
thus to a superalgebra anti-homomorphism: $\underline{S}: P_{B}
\rightarrow P_{B}$, with values on the generators determined by
\eqref{eq:HopfPB}.

Now it is sufficient to verify the rest of the super-Hopf algebra
axioms (coassociativity, counity and the compatibility condition
for the antipode) on the generators of $P_{B}$. This can be done
with straigthforward computations.
\end{proof}
Let us note here, that the above proposition generalizes a result
which -in the case of finite degrees of freedom- is a direct
consequence of the work in \cite{Pal}. In that work the
parabosonic algebra in $2n$ generators ($n$-paraboson algebra)
$P_{B}^{(n)}$ is shown to be isomorphic to the universal
enveloping algebra of the orthosymplectic Lie superalgebra:
$P_{B}^{(n)} \cong U(B(0,n))$. We present this accomplishment in
detail in \seref{3}. See also the discussion in \cite{KaDa2}.

\section{Lie super-algebraic structure of Parabosons:
 the case of finite degrees of freedom}\selabel{3}

In this section, we restrict ourselves to the case of the finite
degrees of freedom (finite number of parabosons), in order to
recall an important development in the study of the structure of
the parabosonic algebra. We thus consider the parabosonic algebra
generated by $B_{i}^{+}, B_{j}^{-}, I, \ $ for $i,j=1, 2, ...n$
where $n$ is a positive integer. The generators satisfy exactly
the same relations as before, determined by \eqref{CCRparabose} or
equivalently \eqref{eq:pbdef}. The difference is that we only have
a finite number of generators now and we will call this algebra
the parabosonic algebra in $2n$ generators or the $n$-paraboson
algebra from now on. We are going to denote it by: $P_{B}^{(n)}$.

It  was conjectured \cite{OmOhKa}, that due to the mixing of
commutators and anticommutators in $P_{B}^{(n)}$ the proper
mathematical ``playground" for the study of the structure of
$P_{B}^{(n)}$ should be some kind of Lie superalgebra (
$\mathbb{Z}_{2}$-graded Lie algebra). Starting in the early '80
's, and using the recent (by that time) results in the
classification of the finite dimensional simple complex Lie
superalgebras which was obtained by Kac (see: \cite{Kac1, Kac2}
but also \cite{kap}), Palev managed to identify the parabosonic
algebra with the universal enveloping algebra of a certain simple
complex Lie superalgebra. In \cite{Pal}, \cite{Pal5} and
\cite{Pal2}, Palev shows the following:
\begin{lemma}
In the k-vector space $P_{B}^{(n)}$ we consider the  k-subspace
generated by the set of elements:
$$
\Big\{ \{B_{i}^{\xi}, B_{j}^{\eta}\}, \  B_{k}^{\epsilon} \ \ |
\xi, \eta, \epsilon = \pm, \ i,j,k = 1,\ldots,n \Big\}
$$
This vector space is a superspace (i.e.: a $\mathbb{Z}_{2}$-graded
vector space or equivalently: a $\mathbb{CZ}_{2}$-module).The
elements $B_{i}^{\xi}$ span the odd subspace and the elements
$\{B_{i}^{\xi}, B_{j}^{\eta}\}$ span the even subspace. \\
 The above vector space endowed with a bilinear multiplication $\langle..,..\rangle$
 whose values are determined by the values of
the anticommutator and the commutator in $P_{B}^{(n)}$, i.e.:
$$
\langle B_{i}^{\xi}, B_{j}^{\eta} \rangle = \{B_{i}^{\xi},
B_{j}^{\eta}\}
$$
and:
$$
\big\langle \{ B_{i}^{\xi},  B_{j}^{\eta}\}, B_{k}^{\epsilon}
\big\rangle = \big[ \{ B_{i}^{\xi},  B_{j}^{\eta}\},
B_{k}^{\epsilon} \big] = (\epsilon - \eta)\delta_{jk}B_{i}^{\xi} +
(\epsilon - \xi)\delta_{ik}B_{j}^{\eta}
$$
respectively, according to the above mentioned gradation, is a
simple, complex Lie superalgebra (or: $\mathbb{Z}_{2}$-graded Lie
algebra) isomorphic to $B(0,n)$. \\
The elements:
$$
\begin{array}{ccccc}
  - \frac{1}{2} \{ B_{i}^{-}, B_{i}^{+} \}, & \{ B_{i}^{-}, B_{j}^{+} \}, & \{ B_{i}^{\xi}, B_{j}^{\xi}
  \},
  & (B_{i}^{\xi})^{2}, & B_{i}^{\xi} \\
\end{array}
$$
for all values $i \neq j = 1, 2, \ldots n$ and $\xi = \pm$ ,
constitute a Cartan-Weyl basis of $B(0,n)$.
\end{lemma}

Note that, according to the above lemma, the elements $\big\{
\{B_{i}^{\xi}, B_{j}^{\eta}\} \ \ | \xi, \eta = \pm, \ i,j =
1,\ldots,n \big\}$ constitute a basis in the even part of
$B(0,n)$. This is a subalgebra of $B(0,n)$ isomorphic to the Lie
algebra $sp(2n)$. It's Lie multiplication can be readily deduced
from the above given commutators and reads:
$$
\begin{array}{c}
 \big\langle \{ B_{i}^{\xi},  B_{j}^{\eta}\}, \{ B_{k}^{\epsilon},
B_{l}^{\phi} \} \big\rangle = \big[ \{ B_{i}^{\xi},
B_{j}^{\eta}\}, \{ B_{k}^{\epsilon}, B_{l}^{\phi} \} \big] =
  \\
      \\
  (\epsilon - \eta)\delta_{jk} \{ B_{i}^{\xi}, B_{l}^{\phi} \} +
(\epsilon - \xi)\delta_{ik}\{ B_{j}^{\eta}, B_{l}^{\phi} \} +
(\phi - \eta)\delta_{jl}\{ B_{i}^{\xi}, B_{k}^{\epsilon} \} +
(\phi - \xi)\delta_{il}\{ B_{j}^{\eta}, B_{k}^{\epsilon} \} \\
\end{array}
$$
On the other hand the elements $\big\{ B_{k}^{\epsilon} \ \ |
\epsilon = \pm, \ k = 1,\ldots,n \big\}$ constitute a basis of the
odd part of $B(0,n)$.

 Note also, that $B(0,n)$  in Kac's
notation, is the classical simple complex orthosymplectic Lie
superalgebra denoted $osp(1,2n)$ in the notation traditionally
used by physicists until then.

Based on the above observations, Palev finally proves:

\begin{proposition} \prlabel{parab}
The parabosonic algebra in $2n$ generators is isomorphic to the
universal enveloping algebra of the classical simple complex Lie
superalgebra $B(0,n)$ (according to the classification of the
simple complex Lie superalgebras given by Kac), i.e:
$$
P_{B}^{(n)} \cong U(B(0,n))
$$
\end{proposition}

Lie superalgebras are exactly the algebraic structures underlying
the idea of supersymmetry. The above mentioned proposition thus,
indicates a link between parafield theories and supersymmetry. For
a similar discussion one should also see \cite{Pl1}.

\prref{parab} also indicates that in the case of the finite
degrees of freedom, the representation theory of the parabosonic
algebra $P_{B}^{(n)}$ coincides with the representation theory of
the orthosymplectic Lie superalgebra  $osp(1/2n)$ \cite{Kac3}.

In the case of the finite degrees of freedom, the super Hopf
structure of the parabosonic algebra $P_{B}^{(n)}$ can be deduced
from the fact that the universal enveloping algebra $U(L)$ of any
Lie superalgebra $L$ is an super Hopf algebra. In the case of the
infinite degrees of freedom, the parabosonic algebra is referred
in the bibliography \cite{Pal1} to be also the universal
enveloping algebra of some Lie superalgebra. Let us stress here
however, that our proof of \prref{superHopfPb} does not make use
of any kind of underlying Lie superalgebraic structure for either
the $P_{B}^{(n)}$ or the $P_{B}$ algebras.


\section{Ordinary Hopf structures in Parabosons}\selabel{4}

\subsection{Review of the bosonisation technique}

A general scheme for transforming a Hopf algebra $A$ in the
braided monoidal category ${}_{H}\mathcal{M}$ (where $H$ is a
quasitriangular Hopf algebra) into an ordinary one, namely the
smash product Hopf algebra $A \star H$, such that the category of
braided modules of $A$ and the category of (ordinary) modules of
$A \star H$ are equivalent, has been developed in the original
reference \cite{Maj1}, see also \cite{Maj2, Maj3, Mon}.  The
technique is called bosonisation, the term coming from physics.
This technique uses ideas developed by Molnar in \cite{Mo} and by
Radford in \cite{Ra}, which guarantee the compatibility between an
algebraic and a coalgebraic structure in a tensor product
\cite{Mo} or even in a smash product \cite{Ra}, in order for it to
become a bialgebra and finally a Hopf algebra. It is also
presented and applied in \cite{Andru, Fi, FiMon}. For clarity
reasons we give a compact review the main points of the above
method.

 In general, $A$ being a Hopf algebra in a category, means that
 $A$ apart from being an algebra and a coalgebra, is also an object of
 the category and at the same time its structure maps (commultiplication, antipode etc)
  are morphisms in the category.
 In particular, if $H$ is some quasitriangular Hopf algebra, $A$ being
a Hopf algebra in the braided monoidal category
${}_{H}\mathcal{M}$ , means that the $H$-module $A$ is an algebra
in ${}_{H}\mathcal{M}$ (or $H$-module algebra) and a coalgebra in
${}_{H}\mathcal{M}$ (or $H$-module coalgebra) and at the same time
$\Delta_{A}$ and $\varepsilon_{A}$ are algebra morphisms in the
category ${}_{H}\mathcal{M}$. (For more details on the above
definitions one may consult for example \cite{Maj2, Maj3} or
\cite{Mon}).

 Since $A$ is an $H$-module algebra we can form the
cross product algebra $A \rtimes H$ (also called: smash product
algebra) which as a k-vector space is $A \otimes H$ (i.e. we
write: $a \rtimes h \equiv a \otimes h$ for every $a \in A$, $h
\in H$), with multiplication given by:
\begin{equation} \label{eq:crosspralg}
(b \otimes h)(c \otimes g) = \sum b(h_{1} \vartriangleright c)
\otimes h_{2}g
\end{equation}
for all $b,c \in A$ and $h,g \in H$, the $\otimes$ the usual
tensor product and $\Delta (h) = \sum h_1 \otimes h_2$.

  On the other hand $A$ is a (left) $H$-module
coalgebra with $H$ becomes qua\-sitriangular through the
$R$-matrix $R_{H} = \sum R_{H}^{(1)} \otimes R_{H}^{(2)}$.
 Quasitriangularity
switches the (left) action of $H$ on $A$ into a (left) coaction
$\rho: A \rightarrow H \otimes A$ through:
\begin{equation} \label{eq:act-coact}
\rho(a) = \sum R_{H}^{(2)} \otimes (R_{H}^{(1)} \vartriangleright
a)
\end{equation}
and $A$ endowed with this coaction becomes (see \cite{Maj2, Maj3})
a (left) $H$-comodule coalgebra or equivalently a coalgebra in
${}^{H}\mathcal{M}$ (meaning that $\Delta_{A}$ and
$\varepsilon_{A}$ are (left) $H$-comodule morphisms, see
\cite{Mon}).

 We recall here (see: \cite{Maj2, Maj3}) that when
$H$ is a Hopf algebra and $A$ is a (left) $H$-comodule coalgebra
with the (left) $H$-coaction given by: $\rho(a) = \sum a^{(1)}
\otimes a^{(0)}$ , one may form the cross coproduct coalgebra $A
\rtimes H$, which as a k-vector space is $A \otimes H$ (i.e. we
write: $a \rtimes h \equiv a \otimes h$ for every $a \in A$, $h
\in H$), with comultiplication given by:
\begin{equation} \label{eq:crosscoprcoalg}
\Delta(a \otimes h) = \sum a_{1} \otimes a_{2}^{ \ (1)} \ h_{1}
\otimes a_{2}^{ \ (0)} \otimes h_{2}
\end{equation}
and counit: $\varepsilon(a \otimes h) = \varepsilon_{A}(a)
\varepsilon_{H}(h)$. (In the above: $\Delta_{A}(a) = \sum a_{1}
\otimes a_{2}$ and we use in the elements of $A$ upper indices
included in parenthesis to denote the components of the coaction
according to the Sweedler notation, with the convention that
$a^{(i)} \in H$
for $i \neq 0$). \\
Now we proceed by applying the above described construction of the
cross coproduct coalgebra $A \rtimes H$ , with the special form of
the (left) coaction given by eq. \eqref{eq:act-coact}. Replacing
thus eq. \eqref{eq:act-coact} into eq. \eqref{eq:crosscoprcoalg}
we get for the special case of the quasitriangular Hopf algebra H
the cross coproduct comultiplication:
\begin{equation} \label{eq:crosscoprcoalgR}
\Delta(a \otimes h) = \sum a_{1} \otimes R_{H}^{(2)}h_{1} \otimes
(R_{H}^{(1)} \vartriangleright a_{2}) \otimes h_{2}
\end{equation}
Finally we can show that the cross product algebra (with
multiplication given by \eqref{eq:crosspralg}) and the cross
coproduct coalgebra (with comultiplication given by
\eqref{eq:crosscoprcoalgR}) fit together and form a bialgebra
(see: \cite{Maj2, Maj3, Mo, Mon, Ra}). This bialgebra, furnished
with an antipode
\begin{equation} \label{antipodecrosspr}
S(a \otimes h) = (S_{H}(h_{2}))u(R^{(1)} \vartriangleright
S_{A}(a)) \otimes S(R^{(2)}h_{1})
\end{equation}
where $u = \sum S_{H}(R^{(2)})R^{(1)}$, and $S_{A}$ the (braided)
antipode of $A$, becomes (see \cite{Maj2}) an ordinary Hopf
algebra. This is the smash product Hopf algebra denoted $A \star
H$.

Apart from the above described construction, it is worth
mentioning two more important points proved in \cite{Maj1}. First,
it is shown that if $H$ is triangular and $A$ is quasitriangular
in the category ${}_{H}\mathcal{M}$, then $A \star H$ is
(ordinarily) quasitriangular.
 Second, it is shown that the braided modules of the original braided Hopf algebra
$A$ ($A$-modules in ${}_{H}\mathcal{M}$, where $A$ is an algebra
in ${}_{H}\mathcal{M}$) and the (ordinary) modules of the
``bosonised" (ordinary) Hopf algebra $A \star H$ are in a
bijective correspondence, providing thus an equivalence of
categories the category of the braided modules of $A$ ($A$-modules
in ${}_{H}\mathcal{M}$) where the braiding is given by a natural
family of isomorphisms $\Psi_{V,W}: V \otimes W \cong W \otimes
V$, stated explicitly by
\begin{equation} \label{eq:braid2}
\Psi_{V,W}(v \otimes w) = \sum (R_{H}^{(2)} \vartriangleright w)
\otimes (R_{H}^{(1)} \vartriangleright v)
\end{equation}
for any $V,W \in obj({}_{H}\mathcal{M})$ (by $v,w$ we denote any
 elements of $V,W$ respectively), is equivalent to the
category of the (ordinary) modules of $A \star H$. Let us stress
here, that from the mathematicians viewpoint, this does not prove
that we have a Morita equivalence, since such a kind of
equivalence would presuppose the whole category of modules over
$A$ and not it's subcategory of braided modules.

Let us close this review of the bosonisation technique, with a
note on terminology. The term ``bosonisation" was first introduced
by Majid in \cite{Maj1}. It is coming from physics and it stems
from the -widespread among physicists- point of view which
considers the bosonic algebra to be a quotient algebra of the
universal enveloping algebra of the Heisenberg Lie algebra, with
its elements thus being even or: ungraded elements.

 In the case that
$H = \mathbb{CG}$ where $\mathbb{G}$ is a finite abelian group,
the Hopf algebra in ${}_{\mathbb{CG}}\mathcal{M}$ is just a
$\mathbb{G}$-graded Hopf algebra in the sense of \cite{Ko},
\cite{Mon} or \cite{Scheu}. The result of the bosonisation
technique in this case is the construction of an ordinary Hopf
algebra $A \star \mathbb{CG}$ which absorbs the grading and whose
elements are ungraded or ``bosonic" elements. This is the original
motivation which led Majid to the use of the term bosonisation
(see also \cite{Maj2, Maj3}).

Finally, let us note that for another use of the term
bosonisation, which is technically reminiscent of the above but it
is not explicitly related to the Hopf structure, one should also
see \cite{Pl2}.

   \subsection{Bosonisation of $P_{B}$ using the smash product}

 In the special case that $A$ is some super-Hopf
algebra, then: $H = \mathbb{CZ}_{2}$, equipped with it's
non-trivial quasitriangular structure, formerly mentioned. In this
case, the technique simplifies and the ordinary Hopf algebra
produced is the smash product Hopf algebra $A \star
\mathbb{CZ}_{2}$. The grading in $A$ is induced by the
$\mathbb{CZ}_{2}$-action on $A$:
\begin{equation} \label{eq:cz2action}
1\vartriangleright a = a,\quad  g \vartriangleright a = (-1)^{|a|}a
\end{equation}
for $a$ homogeneous in $A$. Utilizing the non-trivial $R$-matrix
$R_{g}$ and using eq. \eqref{eq:nontrivRmatrcz2} and eq.
\eqref{eq:act-coact} we can readily deduce the form of the induced
$\mathbb{CZ}_{2}$-coaction on $A$:
\begin{equation} \label{eq:cz2coaction}
\rho(a) = g^{|a|} \otimes a \equiv\left\{ \begin{array}{ccc}
    1 \otimes a & , & a: \textrm{even} \\
    g \otimes a & , & a: \textrm{odd} \\
\end{array} \right.
\end{equation}
Let us note here that instead of invoking the non-trivial
quasitriangular structure $R_{g}$ we could alternatively extract
the (left) coaction \eqref{eq:cz2coaction} utilizing the
self-duality of the $\mathbb{CZ}_{2}$ Hopf algebra. For any
abelian group $\mathbb{G}$ a (left) action of $\mathbb{CG}$
coincides with a (right) action of $\mathbb{CG}$. On the other
hand, for any finite group, a (right) action of $\mathbb{CG}$ is
the same thing as a (left) coaction of the dual Hopf algebra
$(\mathbb{CG})^{*}$. Since $\mathbb{CZ}_{2}$ is both finite and
abelian and hence self-dual in the sense that: $\mathbb{CZ}_{2}
\cong (\mathbb{CZ}_{2})^{*}$ as Hopf algebras, it is immediate to
see that the (left) action \eqref{eq:cz2action} and the (left)
coaction \eqref{eq:cz2coaction} are virtually the same thing.

 The above mentioned action and coaction enable us to form the
cross product algebra and the cross coproduct coalgebra according
to the preceding discussion which finally form the smash product
Hopf algebra $A \star \mathbb{CZ}_{2}$.  The grading of $A$, is
``absorbed" in $A \star \mathbb{CZ}_{2}$, and becomes an inner
automorphism:
$$
gag = (-1)^{|a|}a
$$
where we have identified: $a \star 1 \equiv a$ and $1 \star g
\equiv g$ in $A \star \mathbb{CZ}_{2}$ and $a$ be a homogeneous
element in $A$. This inner automorphism is exactly the adjoint
action of $g$ on $A \star \mathbb{CZ}_{2}$ (as an ordinary Hopf
algebra). The following proposition is proved -as an example of
the bosonisation technique- in \cite{Maj2}:
\begin{proposition} \label{bosonisat}
Corresponding to every super-Hopf algebra $A$ there is an ordinary
Hopf algebra $A \star \mathbb{CZ}_{2}$, its bosonisation,
consisting of $A$ extended by adjoining an element $g$ with
relations, coproduct, counit and antipode:
\begin{equation} \label{eq:HopfPBg}
\begin{array}{cccc}
  g^{2} = 1 & ga = (-1)^{|a|}ag & \Delta(g) = g \otimes g & \Delta(a) = \sum a_{1}g^{|a_{2}|} \otimes a_{2} \\
                                     \\
  S(g) = g & S(a) = g^{-|a|}\underline{S}(a) & \varepsilon(g) = 1 & \varepsilon(a) = \underline{\varepsilon}(a) \\
\end{array}
\end{equation}
where $\underline{S}$ and $\underline{\varepsilon}$ denote the
original maps of the super-Hopf algebra $A$.

 In the case that $A$
is super-quasitriangular via the $R$-matrix $$ \ \underline{R} =
\sum \underline{R}^{(1)} \otimes \underline{R}^{(2)} \ $$ then the
bosonised Hopf algebra $A \star \mathbb{CZ}_{2}$ is
quasitriangular (in the ordinary sense) via the $R$-matrix: $$ \
R_{smash} = R_{Z_{2}} \sum
\underline{R}^{(1)}g^{|\underline{R}^{(2)}|} \otimes
\underline{R}^{(2)} \ $$

 Moreover, the representations of the bosonised Hopf algebra $A
\star \mathbb{CZ}_{2}$ are precisely the super-representations of
the original superalgebra $A$.
\end{proposition}
The application of the above proposition in the case of the
parabosonic algebra $P_{B}$ is straightforward, we immediately get
it's bosonised form $P_{B(g)}$ which by definition is $ P_{B(g)}
\equiv P_{B} \star \mathbb{CZ}_{2} $. Utilizing equations
\eqref{eq:HopfPB} which describe the super-Hopf algebraic
structure of the parabosonic algebra $P_{B}$, and replacing them
into equations \eqref{eq:HopfPBg} which describe the ordinary Hopf
algebra structure of the bosonised superalgebra, we get after
straightforward calculations the explicit form of the (ordinary)
Hopf algebra structure of $P_{B(g)} \equiv P_{B} \star
\mathbb{CZ}_{2}$ which reads:
\begin{equation} \label{eq:HopfPBgexpl}
\begin{array}{cccc}
  \Delta(B_{i}^{\pm}) = B_{i}^{\pm} \otimes 1 + g \otimes B_{i}^{\pm} & \Delta(g) = g \otimes g
  & \varepsilon(B_{i}^{\pm}) = 0 & \varepsilon(g) = 1  \\
          \\
  S(B_{i}^{\pm}) = B_{i}^{\pm}g = -gB_{i}^{\pm} & S(g) = g & g^{2} = 1 & \{g,B_{i}^{\pm}\} = 0  \\
\end{array}
\end{equation}
where $i = 1, 2, \ldots \ $ and we have again identified
$B_{i}^{\pm} \star 1 \equiv B_{i}^{\pm}$ and $1 \star g \equiv g$
in $P_{B} \star \mathbb{CZ}_{2}$.

Finally, we can easily check that since $\mathbb{CZ}_{2}$ is
triangular (via $R_{Z_{2}}$) and $P_{B}$ is super-quasitriangular
(trivially since it is super-cocommutative)
 it is an immediate consequence of the above proposition that
 $P_{B(g)}$ is quasitriangular (in the ordinary sense) via the
 $R$-matrix:
\begin{equation} \label{eq:nontrivRmatrcz2smash}
R_{smash} = \frac{1}{2}(1 \star 1 \otimes 1 \star 1 + 1 \star 1
\otimes 1 \star g + 1 \star g \otimes 1 \star 1 - 1 \star g
\otimes 1 \star g)
\end{equation}
which under the above mentioned identification: $1 \star g \equiv
g$ completely coincides with the $R$-matrix $R_{Z_{2}}$ given in
eq. \eqref{eq:nontrivRmatrcz2}.

\subsection{Bosonisation of $P_{B}$ using two additional operators $K^{\pm}$}

Let us describe now a  different construction (see also:
\cite{DaKa, KaDa2} for the case of the finite degrees of freedom
and \cite{KaDa1} for the general case), which achieves the same
object, i.e. the determination of an ordinary Hopf structure for
the parabosonic algebra $P_{B}$.
\begin{proposition} \label{altern}
Corresponding to the super-Hopf algebra $P_{B}$ there is an
ordinary Hopf algebra $P_{B(K^{\pm})}$, consisting of $P_{B}$
extended by adjoining two elements $K^{+}$, $K^{-}$ with
relations, coproduct, counit and antipode:
\begin{equation} \label{eq:HopfPBK}
\begin{array}{cc}
  \Delta(B_{i}^{\pm}) = B_{i}^{\pm} \otimes 1 + K^{\pm} \otimes B_{i}^{\pm} & \Delta(K^{\pm}) = K^{\pm} \otimes K^{\pm} \\
       \\
  \varepsilon(B_{i}^{\pm}) = 0 & \varepsilon(K^{\pm}) = 1 \\
                     \\
  S(B_{i}^{\pm}) = B_{i}^{\pm}K^{\mp} & S(K^{\pm}) = K^{\mp} \\
                      \\
  K^{+}K^{-} = K^{-}K^{+} = 1 & \{K^{+},B_{i}^{\pm}\} = 0 = \{K^{-},B_{i}^{\pm}\} \\
\end{array}
\end{equation}
for  all values $i = 1, 2, \ldots \ $.
\end{proposition}

\begin{proof}
Consider the complex vector space $\mathbb{C}\langle X_{i}^{+},
X_{j}^{-}, K^{\pm} \rangle$ freely generated by the elements
$X_{i}^{+}, X_{j}^{-}, K^{+}, K^{-}$ where $i = 1, 2, \ldots \ $.
Denote $T(X_{i}^{+}, X_{j}^{-}, K^{\pm})$ its tensor algebra. In
the tensor algebra we denote $I_{BK}$ the ideal generated by all
the elements of the form \eqref{eq:pbdef} together with all
elements of the form: $\ K^{+}K^{-}-1 \ $, $\ K^{-}K^{+}-1 \ $, $\
\{K^{+}, X_{i}^{\pm}\} \ $, $\ \{K^{-}, X_{i}^{\pm}\} \ $, for all
values of $ \ i = 1, 2, \ldots \ $. We define:
$$
P_{B(K^{\pm})} = T(X_{i}^{+}, X_{j}^{-}, K^{\pm}) / I_{BK} \
$$
We denote by $B_{i}^{\pm}, K^{\pm}$ where $i = 1, 2, \ldots \ $
the images of the generators $X_{i}^{\pm}, K^{\pm}$, $ \ i = 1, 2,
\ldots \ $ of the tensor algebra, under the canonical projection.
These are a set of generators of $P_{B(K^{\pm})}$. \\
 Consider the linear map
$$
\Delta^{T} : \mathbb{C}\langle X_{i}^{+}, X_{j}^{-}, K^{\pm}
\rangle \rightarrow P_{B(K^{\pm})} \otimes P_{B(K^{\pm})}
$$
determined by
$$
\begin{array}{c}
\Delta^{T}(X_{i}^{\pm}) = B_{i}^{\pm}
\otimes 1 + K^{\pm} \otimes B_{i}^{\pm} \\
    \\
 \Delta^{T}(K^{\pm}) = K^{\pm} \otimes K^{\pm}  \\
\end{array}
$$
By the universality property of the tensor algebra, this map is
uniquely extended to an algebra homomorphism:
$$
\Delta^{T}: T(X_{i}^{+}, X_{j}^{-}, K^{\pm}) \rightarrow
P_{B(K^{\pm})} \otimes P_{B(K^{\pm})}
$$
We emphasize that the usual tensor product algebra $P_{B(K^{\pm})}
\otimes P_{B(K^{\pm})}$ is now considered, with multiplication $(a
\otimes b)(c \otimes d) = ac \otimes bd$ for any $a,b,c,d \in
P_{B(K^{\pm})}$. Now we can trivially verify that
\begin{equation} \label{eq:DKb}
 \Delta^{T}(\{K^{\pm},X_{i}^{\pm}\})
= \Delta^{T}(K^{+}K^{-} -1) = \Delta^{T}(K^{-}K^{+}-1) = 0
\end{equation}
After lengthy algebraic calculations we also get:
\begin{equation} \label{eq:Db}
\Delta^{T}(\big[ \{ X_{i}^{\xi},  X_{j}^{\eta}\}, X_{k}^{\epsilon}
\big] -
 (\epsilon - \eta)\delta_{jk}X_{i}^{\xi} - (\epsilon -
 \xi)\delta_{ik}X_{j}^{\eta}) = 0
\end{equation}
The calculations are carried in the same spirit of the calculation
found in Appendix \ref{app} but we note that this time we use the
comultiplication stated in equation \eqref{eq:HopfPBK} and the
usual tensor product algebra $P_{B(K^{\pm})} \otimes
P_{B(K^{\pm})}$ is considered instead of the braided tensor
product algebra $P_{B(K^{\pm})} \underline{\otimes}
P_{B(K^{\pm})}$ used in Appendix \ref{app}.

Relations \eqref{eq:DKb}, and \eqref{eq:Db}, mean that $I_{BK}
\subseteq ker (\Delta^{T})$ which in turn implies that
$\Delta^{T}$ is uniquely extended to an algebra homomorphism from
$ P_{B(K^{\pm})}$ to  the usual tensor product algebra
$P_{B(K^{\pm})} \otimes P_{B(K^{\pm})}$, with the values on the
generators determined by \eqref{eq:HopfPBK}, according to the
following (commutative) diagram:
\begin{displaymath}
\xymatrix{T(X_{i}^{+}, X_{j}^{-}, K^{\pm}) \ar[rr]^{\Delta^{T}}
\ar[dr]_{\pi} & & P_{B(K^{\pm})} \otimes
P_{B(K^{\pm})} \\
 & P_{B(K^{\pm})} \ar@{.>}[ur]^{ \ \Delta} & }
\end{displaymath}
 Following the same
procedure we construct an algebra homomorphism $\varepsilon:
P_{B(K^{\pm})} \rightarrow \mathbb{C}$ and an algebra
antihomomorphism $S: P_{B(K^{\pm})} \rightarrow P_{B(K^{\pm})}$
which are completely determined by their values on the generators
of $P_{B(K^{\pm})}$ as given in \eqref{eq:HopfPBK}. Note that in
the case of the antipode we start by defining a linear map $S^{T}$
from $\mathbb{C}\langle X_{i}^{+}, X_{j}^{-}, K^{\pm} \rangle$ to
the opposite algebra $P_{B(K^{\pm})}^{op}$, with values determined
by $S^{T}(X_{i}^{\pm}) = B_{i}^{\pm}K^{\mp}$ and $S^{T}(K^{\pm}) =
K^{\mp} \ $. Following the above described procedure we verify
that $I_{BK} \subseteq ker(S^{T})$, thus resulting with an algebra
anti-homomorphism:
$$
S: P_{B(K^{\pm})} \rightarrow P_{B(K^{\pm})}
$$
with values on the generators determined by \eqref{eq:HopfPBK}.

 Now it is
sufficient to verify the rest of the Hopf algebra axioms (i.e.:
coassociativity of $\Delta$, counity property for $\varepsilon$,
and the compatibility condition which ensures us that $S$ is an
antipode) on the generators of $P_{B(K^{\pm})}$. This can be done
with straightforward computations (see \cite{DaKa}).
\end{proof}
Let us notice here, that the initiation for the above mentioned
construction lies in the case of the finite degrees of freedom: If
we consider the parabosonic algebra in $2n$ generators
($n$-paraboson algebra) and denote it $P_{B}^{(n)}$, it is
possible to construct explicit realizations of the elements
$K^{+}$ and $K^{-}$ in terms of formal power series, such that the
relations specified in \eqref{eq:HopfPBK} hold. The construction
is briefly (see also \cite{DaKa}) as follows: We define
$$
\mathcal{N} = \sum_{i=1}^{n}N_{ii} =
\frac{1}{2}\sum_{i=1}^{n}\{B_{i}^{+},B_{i}^{-}\}
$$
We inductively prove
$$
\mathcal{N}^{m} B_{i}^{\pm}= B_{i}^{\pm}\left(\mathcal{N}\pm I\right)^{m}
$$
For any entire complex function $f(z)$ we get
\[
f\left(\mathcal{N}\right)B_{i}^{\pm} =B_{i}^{\pm}  f\left(\mathcal{N}+I\right)
\]

We now introduce the following elements:
$$
\begin{array}{ccccc}
  K^{+} = \exp(i \pi \mathcal{N}), \quad K^{-} = \exp(-i \pi \mathcal{N}) \\
\end{array}
$$
then we get
\begin{equation} \label{eq:Kb}
\begin{array}{lr}
 \{K^{+},B_{i}^{\pm}\} = 0, &  \{K^{-},B_{i}^{\pm}\} = 0 \\
\end{array}
\end{equation}
A direct application of the Baker-Campbell-Hausdorff formula leads
also to:
\begin{equation} \label{eq:KK}
K^{+}K^{-} = K^{-}K^{+} = 1
\end{equation}
which completes the statement.

\section{Discussion}
Several points which deserve to be discussed stem from the
constructions of the preceding paragraphs:

First of all we should mention that an analogous treatment
regarding the (super-) algebraic and the (super-) Hopf algebraic
structure can be given for the parafermionic algebras and for
mixed systems of paraparticles as well. The parafermionic algebra
in finite degrees of freedom has been shown \cite{Kata},
\cite{RySu}, to be isomorphic to the universal enveloping algebra
of the Lie algebra $B_{n} = so(2n+1)$ and thus an ordinary Hopf
algebra \cite{DaKa}, consequently the grading does not seem to
play an important role in it's structure. On the other hand,
algebras which describe mixed systems of paraparticles such as the
relative parafermi or the relative parabose sets (see
\cite{GreeMe} for their description) have been shown to be
$\mathbb{Z}_{2}$-graded (see \cite{Pal6}) or $\mathbb{Z}_{2}
\times \mathbb{Z}_{2}$-graded (see \cite{Ya}) algebras
respectively. It would thus be an interesting idea to apply
similar techniques to these algebras and obtain results about
their braided representations and their tensor products, and about
their super-Hopf structure and their bosonised forms as well. Of
course such questions inevitably involve questions of pure
mathematical interest, such as the possible quasitriangular
structures (and thus the possible braidings) for a
$\mathbb{C}(\mathbb{Z}_{2} \times \mathbb{Z}_{2})$ group Hopf
algebra, which up to our knowledge have not yet been solved in
general (see \cite{Scheu1} for a
relevant discussion).  \\
Let us note here that the super-Hopf algebraic structure of the
parabosonic algebra established in \seref{2} has important
applications in physics: It has recently been shown \cite{KaDa3}
that using the results of \prref{superHopfPb}, one may obtain the
construction of the parabosonic Fock-like representations
corresponding to an arbitrary value of the positive integer $p$
(see: \cite{GreeMe}) as irreducible submodules of the braided
tensor product representations between $p$-copies of the first
Fock-like representation (corresponding to the value of $p=1$).
The super-Hopf algebraic structure of the parabosonic algebra is
essential in this process and leads us to a purely braided
interpretation of the Green ansatz for parabosons (see
\cite{KaDa3} for a more detailed description of the method).

Regarding now the results of the last section, i.e.: the
``bosonised" variants $P_{B(g)}$, $P_{B(K^{\pm})}$ of the
parabosonic algebras, various questions can be posed: \\
 From the point of view of the structure,
an obvious question arises: While $P_{B(g)}$ is a quasitriangular
Hopf algebra through the $R$-matrix: $R_{Z_{2}}$ given in eq.
\eqref{eq:nontrivRmatrcz2}, there is yet no suitable $R$-matrix
for the Hopf algebra $P_{B(K^{\pm})}$. Thus the question of the
quasitriangular structure of $P_{B(K^{\pm})}$ is open. \\
On the other hand, regarding representations, we have already
noted that the super representations of $P_{B}$ (
$\mathbb{Z}_{2}$-graded modules of $P_{B}$ or equivalently:
$P_{B}$-modules in
 ${}_{\mathbb{CZ}_{2}}\mathcal{M}$ ) are in $``1-1"$ correspodence
 with the (ordinary) representations of $P_{B(g)}$. The construction
 of the representation of $P_{B(g)}$ which
 corresponds to any given representation of $P_{B}$ can be
 done straightforwardly \cite{Maj1, Maj2}. Although we do
 not have such a strong result for the representations of
 $P_{B(K^{\pm})}$, the construction in the end of \seref{4} for the case of
 finite degrees of freedom, enables us to uniquely extend the
 Fock-like \cite{GreeMe} representations of $P_{B}^{(n)}$ to representations of
 $P_{B(K^{\pm})}^{(n)}$. Since the Fock-like representations of
 $P_{B}$ are unique up to unitary equivalence (see the proof in
 \cite{GreeMe} or \cite{OhKa}), this is a point which deserves to be discussed
 analytically in a forthcoming work. We must note here that this
 question has to be discussed in connection with the explicit
 construction of the parabosonic Fock-like representations which
 is yet another open problem (see the discussion in \cite{KaDa3}
 or \cite{LiStVdJ}).

Finally, it will be an interesting thing to study the (ordinary)
tensor products of representations of $P_{B(g)}$ and
$P_{B(K^{\pm})}$, through the comultiplications stated in
\eqref{eq:HopfPBgexpl} and \eqref{eq:HopfPBK} respectively, in
comparison with the (braided) tensor products of (braided)
representations of $P_{B}$ through the comultiplication stated in
\eqref{eq:HopfPB}. Specifically, it will be of interest to answer
the question of whether the ordinary Hopf structures presented in
the last section of this paper are able of generating essentially
new representations of the parabosonic algebra: The possibility
that the reduction of (ordinary) tensor product representations of
either $P_{B(g)}$ or $P_{B(K^{\pm})}$ might lead to submodules
non-equivalent to the parabosonic Fock-like representations (the
latter emerge as irreducible submodules in the reduction of the
braided tensor product representations of $P_{B}$) is an
intriguing one and deserves to be discussed analytically in a
forthcoming work.

\textbf{Acknowledgements:} This paper is part of a
 project supported by ``Pythagoras II", contract number 80897.

\appendix
\appendixpage

\section{Proof of equation (\ref{eq:comultinsuperparab})}\label{app}
Using the fact that the generators of the parabosonic algebra
$P_{B}$ are odd elements and the multiplication in the braided
tensor product algebra $P_{B} \underline{\otimes} P_{B}$ is given
by \eqref{braidedtenspr}, we have: $(I \otimes
B_{i}^{\xi})(B_{j}^{\eta} \otimes I) = - B_{j}^{\eta} \otimes
B_{i}^{\xi}$ while: $(B_{j}^{\eta} \otimes I)(I \otimes
B_{i}^{\xi}) = B_{j}^{\eta} \otimes B_{i}^{\xi}$ in $P_{B}
\underline{\otimes} P_{B}$. Now we compute:
$$
\begin{array}{c}
  \underline{\Delta}^{T}(\{ X_{i}^{\xi}, X_{j}^{\eta}\}) = \underline{\Delta}^{T}(X_{i}^{\xi}X_{j}^{\eta} + X_{j}^{\eta}X_{i}^{\xi})
  = \underline{\Delta}^{T}(X_{i}^{\xi})\underline{\Delta}^{T}(X_{j}^{\eta}) + \underline{\Delta}^{T}(X_{j}^{\eta})\underline{\Delta}^{T}(X_{i}^{\xi}) = \\
      \\
  (B_{i}^{\xi} \otimes I + I \otimes B_{i}^{\xi})(B_{j}^{\eta} \otimes I + I \otimes B_{j}^{\eta}) +  (B_{j}^{\eta} \otimes I + I \otimes B_{j}^{\eta})(B_{i}^{\xi} \otimes I + I \otimes B_{i}^{\xi}) =    \\
    \\
  B_{i}^{\xi}B_{j}^{\eta} \otimes I + B_{i}^{\xi} \otimes B_{j}^{\eta} - B_{j}^{\eta} \otimes B_{i}^{\xi} + I \otimes B_{i}^{\xi}B_{j}^{\eta} +
  +  B_{j}^{\eta}B_{i}^{\xi} \otimes I + \\ \\ + B_{j}^{\eta} \otimes B_{i}^{\xi} - B_{i}^{\xi} \otimes B_{j}^{\eta} + I \otimes B_{j}^{\eta}B_{i}^{\xi}
   = I \otimes \{ B_{i}^{\xi}, B_{j}^{\eta}\} + \{ B_{i}^{\xi}, B_{j}^{\eta}\} \otimes I \\
\end{array}
$$
So we have proved that for the even elements $\{ X_{i}^{\xi},
X_{j}^{\eta}\}$ (for all values of $\xi, \eta, = \pm 1$ and
$i,j=1, 2, ... \ $) of the tensor algebra $T(V_{X})$ we have:
\begin{equation} \label{eq:eventens}
\underline{\Delta}^{T}(\{ X_{i}^{\xi}, X_{j}^{\eta}\}) = \{
B_{i}^{\xi}, B_{j}^{\eta}\} \otimes I + I \otimes \{ B_{i}^{\xi},
B_{j}^{\eta}\}
\end{equation}
Using result \eqref{eq:eventens} and the fact that $\{
X_{i}^{\xi}, X_{j}^{\eta}\}$ (for all values of $\xi, \eta, = \pm
1$ and $i,j=1, 2, ... \ $) are even elements, we get:
$$
\begin{array}{c}
  \underline{\Delta}^{T}(\big[ \{ X_{i}^{\xi},  X_{j}^{\eta}\}, X_{k}^{\epsilon}  \big]) =
  \underline{\Delta}^{T}(\{ X_{i}^{\xi},  X_{j}^{\eta}\}) \underline{\Delta}^{T}(X_{k}^{\epsilon}) -
  \underline{\Delta}^{T}(X_{k}^{\epsilon}) \underline{\Delta}^{T}(\{ X_{i}^{\xi},  X_{j}^{\eta}\}) = \\
     \\
  (\{ B_{i}^{\xi}, B_{j}^{\eta}\} \otimes I + I \otimes \{ B_{i}^{\xi}, B_{j}^{\eta}\})
  (B_{k}^{\epsilon} \otimes I + I \otimes B_{k}^{\epsilon}) \\ \\ - (B_{k}^{\epsilon} \otimes I + I \otimes B_{k}^{\epsilon})
   (\{ B_{i}^{\xi}, B_{j}^{\eta}\} \otimes I + I \otimes \{ B_{i}^{\xi}, B_{j}^{\eta}\}) = \\
    \\
  \{ B_{i}^{\xi}, B_{j}^{\eta}\}B_{k}^{\epsilon} \otimes I + \{ B_{i}^{\xi}, B_{j}^{\eta}\} \otimes B_{k}^{\epsilon} +
  B_{k}^{\epsilon} \otimes \{ B_{i}^{\xi}, B_{j}^{\eta}\} + I \otimes \{ B_{i}^{\xi}, B_{j}^{\eta}\}B_{k}^{\epsilon}  \\
    \\
  - B_{k}^{\epsilon} \{ B_{i}^{\xi}, B_{j}^{\eta}\} \otimes I - B_{k}^{\epsilon} \otimes \{ B_{i}^{\xi}, B_{j}^{\eta}\} -
   \{ B_{i}^{\xi}, B_{j}^{\eta}\} \otimes B_{k}^{\epsilon} - I \otimes B_{k}^{\epsilon} \{ B_{i}^{\xi}, B_{j}^{\eta}\} = \\
    \\
  \big[ \{ B_{i}^{\xi},  B_{j}^{\eta}\}, B_{k}^{\epsilon}  \big] \otimes I + I \otimes
  \big[ \{ B_{i}^{\xi},  B_{j}^{\eta}\}, B_{k}^{\epsilon}  \big] = \\
    \\
  ((\epsilon - \eta)\delta_{jk}B_{i}^{\xi} + (\epsilon -
 \xi)\delta_{ik}B_{j}^{\eta})) \otimes I + I \otimes ((\epsilon - \eta)\delta_{jk}B_{i}^{\xi} + (\epsilon -
 \xi)\delta_{ik}B_{j}^{\eta})) =  \\ \\
   (\epsilon - \eta)\delta_{jk}\underline{\Delta}^{T}(X_{i}^{\xi}) + (\epsilon -
 \xi)\delta_{ik}\underline{\Delta}^{T}(X_{j}^{\eta}) =
 \underline{\Delta}^{T}((\epsilon - \eta)\delta_{jk}X_{i}^{\xi} - (\epsilon -
 \xi)\delta_{ik}X_{j}^{\eta}))
\end{array}
$$
which finally completes the proof.

\end{document}